\documentstyle[12pt]{article}

\title{\begin{flushright}
{\normalsize NUC-MINN-95/25-T\\
TPI-MINN-95/11-T\\
HEP-MINN-TH-1342\\
hep-ph/9602405\\
December 1995 \\}
\end{flushright}
\vspace*{0.3in}
{\bf CHIRAL SYMMETRY \\
AT FINITE TEMPERATURE:\\
LINEAR VS NONLINEAR $\sigma$-MODELS}}
\author{{\bf Alexander Bochkarev}$^{\dagger}$\\
 {\it Theoretical Physics Institute}\\
  {\it University of Minnesota}\\ \vspace*{0.2in} {\it Minneapolis, MN 55455}\\
{\bf Joseph Kapusta}\\
  {\it School of Physics and Astronomy}\\
   {\it University of Minnesota}\\ {\it Minneapolis, MN 55455}}

\date{}

\begin{document}

\maketitle

\begin{center}
Abstract
\end{center}

\noindent
The linear O($N$) sigma model undergoes a symmetry restoring phase
transition at finite temperature.  We show that the nonlinear
O($N$) sigma model also undergoes a symmetry restoring phase transition;
the critical temperatures are the same when the linear model is treated
in mean field approximation and the nonlinear model is
treated to leading plus subleading order in the 1/$N$ expansion.
We also carefully define and study the behavior of $f_{\pi}$ and the
scalar condensate at low temperatures in both models, showing that they
are independent of field redefinition.\\


\noindent $^{\dagger}$ On leave from: {\it INR,
Russian Academy of Sciences, Moscow 117312, Russia}
\vfill \eject

\section{Introduction}

The O($N$) model as a quantum field theory in $d$+1 dimensions
\cite{Wilson} is a basis or prototype for many interesting physical
systems.  The bosonic field ${\bf \Phi}$ has $N$ components.  When the
Lagrangian is such that the vacuum state exhibits spontaneous
symmetry breaking it is known as a sigma model.  This is the case
of interest to us here.  In $d$ = 3 space dimensions the linear
sigma model has the potential
\begin{displaymath}
\frac{\lambda}{4}\left({\bf \Phi}^2 - f_{\pi}^2\right)^2
\end{displaymath}
where $\lambda$ is a positive coupling constant and $f_{\pi}$ is
the pion decay constant.  The model is renormalizable.  In the
limit that $\lambda \rightarrow \infty$ the potential goes over
to a delta function constraint on the length of the field vector
and is then known as a nonlinear sigma model.

The classical limit of the field theory is obtained by neglecting
or freezing out the time variable, leaving a field theory in
$d$ dimensions.  In this limit only the zero Matsubara frequency
of the full $d$ + 1 dimensional theory contributes to the partition
function, and the temperature acts like a coupling constant.
One then has a description of an O($N$) Heisenberg magnet in
$d$ dimensions which is a model for real material systems.
This subject has a vast literature \cite{Ma,Amit}.

When $N$ = 4 one has a model for the low energy dynamics of
Quantum Chromodynamics (QCD).  More explicitly, it is essentially
the unique description of the dynamics of very soft pions.
This is basically due to the isomorphism between the groups
O(4) and SU(2)$\times$SU(2), the latter being the appropriate
group for two flavors of massless quarks in QCD.  The linear sigma
model, including the nucleon, goes back to the work of Gell-Mann
and Levy \cite{GM}.  This subject also has a vast literature.
In the last decade much work has been done on {\em chiral perturbation
theory} which starts with the nonlinear sigma model and adds
higher order, nonrenormalizable, terms to the Lagrangian, ordered
by the dimensionality of the coefficients or field derivatives
\cite{cpt}.  This whole program really has its origins in the classic
works of Weinberg \cite{W1,W2}.

Finally, the standard model of the electroweak interactions, due
to Weinberg, Salam, and Glashow has an SU(2) doublet scalar
Higgs field responsible for spontaneous symmetry breaking.
If one neglects spin-1 gauge fields the Higgs sector is also
an O(4) field theory.

All of these limits are interesting to study at finite temperature.
Magnetic materials typically undergo a phase transition from
an ordered to a disordered state.  If quarks are massless, QCD is
expected to undergo a chiral symmetry restoring phase transition
\cite{PW,Wilczek}.  This may have implications for high
energy nucleus-nucleus collisions; see especially \cite{Rob,Raja}
in this respect.  The electroweak theory is expected to have a
symmetry restoring phase transition, too, at which point the baryon
number of the early universe would have been finally determined
\cite{Sintra}.

The linear sigma model was studied
in the classic papers on relativistic quantum field theories at
finite temperature \cite{Kir,Dolan,W3}.  The usefulness of leaving
$N$ as a parameter arises from the fact that there is only one
other parameter in the problem, the quartic coupling constant
$\lambda$ ($f_{\pi}$ just sets the scale and is held fixed in
our considerations).  For QCD at least, and perhaps for electroweak
theory too (this is not known, it is related to the Higgs mass)
the appropriate limit seems to be $\lambda \gg 1$, possibly
even infinity.  This limit is the nonlinear sigma model.  The
only proposed expansion parameter for this model is 1/$N$.
For $N$ = 4 the first few terms in this expansion may not be
quantitatively reliable but it is a good start.  At least in
the theoretical world we can imagine $N$ as large as we wish.
Presumably the physics does not change qualitatively with $N$
as long as it is greater than one.

Our basic physical interest in this paper is QCD.  Among the
questions that are routinely asked are: Does QCD have a finite
temperature phase transition?  If so, is it associated with
color deconfinement, or with the restoration of the spontaneously
broken chiral symmetry, or are they inextricably intertwined?
What would be the order of this phase transition?  What would
be its critical temperature?  How do pions, the Goldstone bosons
of QCD, decouple as the temperature is raised?  How do the
quark and gluon condensates behave as functions of temperature?
Restricting our attention to $N_f$ flavors of massless quarks,
there are strong arguments \cite{PW,Wilczek} and numerical
lattice computations \cite{lattice} which say that for $N_f = 2$
there is a second order phase transition and for $N_f \ge 3$
there is a first order phase transition.  Lattice calculations
also suggest strongly that chiral symmetry is restored and
color is deconfined at the same temperature \cite{Boyd}.

The QCD Lagrangian is invariant under U($N_f$)$\times$U($N_f$)
transformations.  This is isomorphic to
SU$_L$($N_f$)$\times$SU$_R$($N_f$)$\times$U$_{\rm baryon}$
(1)$\times$U$_{\rm axial}$(1); that is, left- and right-handed
chirality transformations of the quark fields, baryon number
conservation, and the famous axial U(1) symmetry.  The axial
U(1) is broken by quantum effects, particularly instantons.
Baryons can be added to the sigma model if desired, but we
shall not do so.  The O($N$) sigma model has no vestige
of the axial U(1), although there do exist versions of the
sigma model based on other groups in which it can be
incorporated.  In general SU($N_f$)$\times$SU($N_f$) is
isomorphic to O($N_f^2$) only for $N_f = 2$.  Otherwise
the interactions are different.  This restricts
any potential quantitative results of our analysis of
the O($N$) model to QCD to $N = 4$.  The very
interesting issue of whether the axial U(1) symmetry is
restored at high temperatures or not cannot be
addressed here \cite{shur2}.

One must be careful in understanding how sigma models are being
applied to the study of QCD at finite temperature.  At very
low temperatures one may argue that the only degrees of freedom
which are excited are pions.  One may then use chiral perturbation
theory to study the thermodynamic properties, as in the classic
works of Leutwyler \cite{Lclassic}.  At the very lowest energies this
is just the nonlinear sigma model.  Note that in this domain one
is really studying a full 3+1 dimensional quantum field theory.
On the other hand, near the critical temperature one may argue
\cite{PW,Wilczek} that the soft long wavelength modes of QCD are in
the same universality class as the O(4) Heisenberg magnet in 3 spatial
directions.  Now one is studying a classical field theory.  The
parameters of the effective free energy functional near the
critical temperature may not be simply related to the parameters
of the sigma field theory model at very low temperature.
Support for this point of view comes from lattice QCD computations
which show that critical exponents are consistent with those of
the O(4) Heisenberg magnet in 3 spatial dimensions \cite{Karsch}.
Conversely, it has been pointed out that in certain models with
composite bosons these arguments might break down \cite{Kogut}
because the compositeness is important near a second order
phase transition.  A particularly useful observation is that
long range correlation functions may be dominated by the soft
modes (pions, sigma meson, ...) near the critical temperature
but the equation of state itself is not \cite{Dine,shur2};
it is dominated by the myriad of other degrees of freedom
(Hagedorn or Particle Data Book compendium of mesons and
baryons or all of the colored quarks and gluons).

We would like to shed some light on just a few of the issues
relating to the above discussion.  First, from the outset
we restrict our attention to the linear and nonlinear sigma
models in 3+1 dimensions.  Within these confines we address
three specific topics: the existence and nature of a chiral
symmetry restoring phase transition, the low temperature dependence
of the ``pion decay constant", and the low temperature dependence
of the ``quark condensate".  The first of these involves
folklore, and the answer, at least in the nonlinear sigma model,
is either obvious to the reader or else very surprising.  For the
latter two we point out some popular misconceptions and reproduce
the existing results (when $N$ = 4) while showing that they are
invariant under field redefinition.

First consider the linear sigma model.  At zero temperature the
effective potential has a shape similar to the bottom of a
wine bottle.  It is minimized by a nonzero value of the field;
this is the condensate.  As the temperature is increased, the radius
of bottom of the potential shrinks, and goes to zero at a
critical temperature of $T_c = \sqrt{12/(N+2)} \, f_{\pi}$.
It is a second order symmetry restoring phase transition.
In the nonlinear sigma model $|{\bf \Phi}|^2$
is fixed at the value $f_{\pi}^2$.  Therefore, it would seem, chiral
symmetry breaking is built into the Lagrangian and there is no
possibility of restoring it at finite temperature.  Another way
of saying this is that there is no order parameter which can go
to zero at finite temperature.  At least this is the folklore in
much of the nuclear and particle physics community!  On the other
hand, the critical temperature in the linear model is independent
of $\lambda$ in the mean field approximation, so one can take
the limit $\lambda \rightarrow \infty$ and still have a phase
transition!  The counter argument to this is that the phase
transition can go away in the limit and so nothing special happens
at the aforementioned value of $T_c$ in the nonlinear model.  We shall
study the nonlinear model directly in section 2 at finite temperature
in the large $N$ approximation.  To leading order in $N$ we shall
show quite straightforwardly that despite the constraint the
nonlinear model has a second order phase transition at a critical
temperature equal to that of the linear model.  We shall show
that this persists to the next to leading order in $N$, although
here we must make an additional high energy approximation.  The
order parameter is identified as is the nature of the two phases.

Next we study what is meant by ``the pion decay constant on finite
temperature".  At zero temperature a common definition is
\begin{equation}
\langle 0| {\cal A}_{\mu}^a |\pi^b(p) \rangle
= i f_{\pi} p_{\mu} \delta^{ab}
\end{equation}
which relates it to the matrix element of the axial vector current
of QCD between the vacuum state and a one pion state of momentum
$p$.  It is difficult, though perhaps not impossible, to generalize
this definition to finite temperature.  Within the linear sigma model
one sometimes sees in the literature $f_{\pi}(T)$ identified with
the thermal average of the sigma field, $v = \langle \sigma \rangle
= |\langle {\bf \Phi} \rangle |$, which is the radius of the bottom
of the effective potential.  In the nonlinear sigma model this radius
is necessarily fixed at $f_{\pi}$.  How then can one understand the
result of Gasser and Leutwyler \cite{Gasser2}
\begin{equation}
f_{\pi}(T) = f_{\pi} \left[ 1 - \frac{N_f}{2}
\left( \frac{T^2}{12 f_{\pi}^2} \right) + \cdots \right]
\label{Gasfpi}
\end{equation}
which was obtained at low temperature in chiral perturbation theory?
(At this order chiral perturbation theory and the nonlinear sigma
model are the same.)  This issue is addressed carefully in section 3.

Finally, a quantity of much interest, especially for the application of
QCD sum rules at finite temperature, is the temperature dependence
of the quark condensate.  Gasser and Leutwyler \cite{Gasser2} and
also Gerber and Leutwyler \cite{Gerber} computed this
quantity at low temperature to be
\begin{equation}
\langle \bar{q} q \rangle = \langle 0 | \bar{q} q | 0 \rangle
\left[ 1 - \frac{N_f^2 - 1}{N_f} \left( \frac{T^2}{12 f_{\pi}^2}
\right) - \frac{N_f^2 - 1}{2N_f^2} \left( \frac{T^2}{12 f_{\pi}^2}
\right)^2 + \cdots \right] \, .
\label{qcon}
\end{equation}
This is obviously a different temperature dependence than that of
$f_{\pi}(T)$.  In addition, one wonders how it is possible to
obtain information on quark condensates from a theory which has
no explicit reference to quarks?  These and related topics are
studied in section 4.

Before beginning the technical part we remark that we will
deal with vacuum divergences in a simple way: we ignore them.
That is, we will always drop temperature independent divergent
quantities from our expressions.  The linear sigma model is
renormalizable and this neglect can easily be rectified; it
does not change any of the principle results.  The nonlinear
model is not renormalizable, although one may consider it to be
the $\lambda \rightarrow \infty$ limit of the linear model.
The short distance physics of these models in the
context of QCD is not correct anyway.

Put another way, it is known that the partition function can be
expressed in terms of the vacuum scattering amplitudes, or S-matrix
elements, for arbitrary reactions involving n particles going in
and m particles coming out \cite{DMB}.  This is the relativistic
virial expansion.  Consider, for example, a real scalar field with a
quartic interaction \cite{TPI}.
The one loop contribution to the partition function is just the
free Bose gas expression.  The two loop contribution corresponds
to two particle scattering with the amplitude evaluated at the
tree level.  The three loop contribution corresponds to two particles
in and four particles out, plus three particles in and three particles
out, plus four particles in and two particles out, with all scattering
amplitudes evaluated at tree level.  The three loop contribution also
has a part which corresponds to a vacuum one loop correction to the
two particle scattering amplitude.  By dropping all vacuum loop
divergencies we are doing a virial expansion with the S matrix
evaluated at the tree level.

An early systematic study of various approximation schemes for
the linear sigma model at finite temperature is \cite{Baym}.
An overall introduction to relativistic quantum field theory
at finite temperature is \cite{JK}.  A somewhat analogous study
of the lattice $O(N)$ Heisenberg model is \cite{Muller}.

\section{Chiral Phase Transition}

It is well known that the linear O($N$) sigma model in 3+1 dimensions
has a second order phase transition when treated in the mean field
approximation.  We briefly repeat that analysis here as a warmup
and precursor to the study of the nonlinear sigma model which is
not nearly as well studied.  We must emphasize that the direct
quantitative applicability to QCD is limited by at least two
factors.  The first is: These sigma models do not have quark
and gluon degrees of freedom and so one can never describe high
temperature quark-gluon plasma with them.  In addition, the compositeness
of the bosons may even influence the phase transition itself
\cite{Kogut}.  The second is: The group SU($N_f$)$\times$SU($N_f$)
is isomorphic to O($N_f^2$) only for $N_f = 2$ and this limits
the analogy to two flavors of massless quarks.  Indeed, lattice
computations of the nonlinear sigma model in 3 dimensions based
on other groups show first order behavior \cite{Dreher}.

The conventions and notation used here are consistent with those
of \cite{JK}.

\subsection{Linear sigma model}

The linear sigma model Lagrangian is
\begin{equation}
{\cal L} = \frac{1}{2}\left(\partial_{\mu}{\bf \Phi}\right)^2
- \frac{\lambda}{4}\left({\bf \Phi}^2 - f_{\pi}^2\right)^2
\end{equation}
where $\lambda$ is a positive coupling constant.  The bosonic field
${\bf \Phi}$ has $N$ components.  Rather arbitarily we define the first
$N - 1$ components to represent a pion field $\mbox{\boldmath $\pi$}$
and the last, $N$'th, component to represent the sigma field.
Since the $O(N)$ symmetry is broken to an $O(N - 1)$ symmetry
at low temperatures, we immediately allow for a sigma condensate
$v$ whose value is temperature--dependent and yet to be determined.
We write
\begin{eqnarray}
\Phi_i({\bf x},t)&=& \pi_i({\bf x},t)
\,\,\,\,\,\,\,\,i = 1,...N-1 \nonumber \\
\Phi_N({\bf x},t)&=& v + \sigma({\bf x},t) \, .
\end{eqnarray}
In terms of these fields the Lagrangian is
\begin{equation}
{\cal L} = \frac{1}{2}\left(\partial_{\mu}\mbox{\boldmath $\pi$}\right)^2
+ \frac{1}{2}\left(\partial_{\mu}\sigma\right)^2
-\frac{\lambda}{4}\left(v^2-f_{\pi}^2+2v\sigma+\sigma^2+
\mbox{\boldmath $\pi$}^2
\right)^2 \, .
\end{equation}
The action at finite temperature is obtained by rotating to imaginary
time, $\tau = i t$, and integrating $\tau$ from 0 to $\beta = 1/T$.
(However, we keep the Minkowski metric; hence, $\partial_{\mu}
= \partial/\partial x^{\mu}$ with $\partial_0 = \partial/\partial t
= i \partial/\partial \tau$.)  The action is defined as
\begin{eqnarray}
S &=& - \frac{\lambda}{4} \left(f_{\pi}^2-v^2\right)^2 \beta V +
\int_0^{\beta} d\tau \int_V d^3x \left\{ \frac{1}{2} \left[
\left(\partial_{\mu}\mbox{\boldmath $\pi$}\right)^2
- \bar{m}_{\pi}^2 \mbox{\boldmath $\pi$}^2
+ \left(\partial_{\mu}\sigma\right)^2 -\bar{m}_{\sigma}^2 \sigma^2 \right]
\right. \nonumber \\
&+& \frac{\lambda}{2} v \left( v^2 - f_{\pi}^2 \right) \sigma
-\left. \lambda v \sigma \left( \mbox{\boldmath $\pi$}^2 + \sigma^2 \right)
-\frac{\lambda}{4}\left( \sigma^2 + \mbox{\boldmath $\pi$}^2
\right)^2 \right\} \, , \label{S}
\end{eqnarray}
where the effective masses are
\begin{eqnarray}
\bar{m}_{\pi}^2&=&\lambda\left(v^2-f_{\pi}^2\right) \nonumber \\
\bar{m}_{\sigma}^2&=&\lambda\left(3v^2-f_{\pi}^2\right) \, .
\label{mass1}
\end{eqnarray}
At any temperature $v$ is chosen such that $\langle \sigma
\rangle = 0$.  This eliminates any one particle reducible (1PR)
diagrams in perturbation theory, leaving only one particle
irreducible (1PI) diagrams.

At zero temperature the potential is minimized when $v = f_{\pi}$.
The pion is massless and the sigma particle has a mass of
$\sqrt{2\lambda} f_{\pi}$.  The Goldstone theorem is satisfied.
Lin and Serot \cite{Lin} have argued that the sigma meson should not be
identified with the attractive s-wave interaction in the $\pi - \pi$
interaction, which is responsible for nuclear attraction.  Rather,
they argue that the sigma meson should have a mass which is at
least 1 GeV if not more.  This means that $\lambda$ is on the order
of 50 or greater.

The simplest approximation at finite temperature is the mean field
approximation.  One allows for $v$ to be temperature dependent;
hence the effective masses are temperature dependent as well.
However, interactions among the particles or collective excitations
are neglected.  The pressure includes only the contribution of
the condensate and of the thermal motion of the independently
moving particles.  Thus
\begin{equation}
P = \frac{T}{V} \ln Z =
- \frac{\lambda}{4} \left(f_{\pi}^2-v^2\right)^2
+ P_0(T,m_{\sigma}) + (N-1) P_0(T,m_{\pi}) \, .
\end{equation}
The pressure of a free relativistic boson gas can be written
several ways:
\begin{equation}
P_0 = -T \int \frac{d^3p}{(2\pi)^3} \ln\left( 1 - {\rm e}^{-\beta
\omega} \right) = \int \frac{d^3p}{(2\pi)^3} \,
\frac{p^2}{3\omega} \, \frac{1}{{\rm e}^{\beta \omega} - 1} \, .
\end{equation}
This is a relatively simple but surprisingly powerful first
approximation which allows one to gain much insight into the
behavior of relativistic quantum field theories at high temperature.
It was used in all the pioneering papers.

One expects that as the temperature is raised, thermal fluctuations
will tend to disorder the condensate field $v$, and at sufficiently
high temperature it may even disappear.  If there is a second order
phase transition then the correlation length should go to infinity,
which is equivalent to the effective sigma mass going to zero.
With such an expectation one may expand the free boson gas pressure
about zero mass to obtain
\begin{equation}
P_0(T,m) = \frac{\pi^2}{90}T^4 - \frac{m^2T^2}{24}
 +\frac{m^3T}{12\pi} + \cdots \, .
\end{equation}
Since the masses are proportional to the square root of $\lambda$
it is generally inconsistent to retain the cubic term in $m$
because there exist loop diagrams which are not included
in the mean field approximation but which
contribute to the same order in $\lambda$.  Therefore we take
\begin{equation}
P(T,v) = N \frac{\pi^2}{90}T^4 + \frac{\lambda}{2}v^2 \left[
f_{\pi}^2 - \frac{N+2}{12}T^2\right] - \frac{\lambda}{4}v^4 \, ,
\end{equation}
where the pion and sigma masses have been expressed in terms
of $\lambda$, $v$ and $f_{\pi}$.  Maximizing the pressure with
respect to $v$ gives
\begin{equation}
v^2 = f_{\pi}^2 - \frac{N+2}{12} T^2 \, .
\label{v}
\end{equation}
This result is easily understood.  Going back to Eq. (\ref{S})
we can differentiate $\ln Z$ with respect to $v$ with the
result that
\begin{equation}
v^2 = f_{\pi}^2 - 3\langle \sigma^2 \rangle - \langle
\pi^2 \rangle
\end{equation}
as long as we choose $\langle \sigma \rangle = 0$.  For any
free bosonic field $\phi$ with mass $m$
\begin{equation}
\langle \phi^2 \rangle =
\int \frac{d^3p}{(2\pi)^3} \, \frac{1}{\omega} \,
\frac{1}{{\rm e}^{\beta \omega} - 1}
\end{equation}
where $\omega = \sqrt{p^2 + m^2}$.  In the limit that the
temperature is greater than the mass $\langle \phi^2 \rangle
\rightarrow T^2/12$.  This yields directly Eq. (\ref{v}).

The condensate goes to zero at a critical temperature given by
\begin{equation}
T_c^2 = \frac{12}{N+2}f_{\pi}^2 \, .
\end{equation}
Above this temperature thermal fluctuations are too large to allow
a nonzero condensate.  It is a straightforward exercise to show
that the pressure and its first derivative is continuous at
$T_{\rm c}$ but that the second derivative is discontinuous.
This is therefore a second order phase transition.

There are two major problems with the mean field approximation
as described.  The first is that the pion has a negative mass
squared at every temperature greater than zero.  Not only is the
Goldstone theorem not satisfied, but there are tachyons as well!
The sigma particle also gets a negative mass squared at temperatures
above $\sqrt{8/(N+2)} \, f_{\pi} < T_{\rm c}$.  This violation of
basic physical principles is resolved by recognizing that the
finite temperature corrections to the squared masses are proportional
to $\lambda T^2$, and that one loop self-energy corrections, not
included in the mean field analysis, are of the same order.
This can be understood with the following analysis.

At high temperatures, when the masses can be neglected in the
loops, the mean field result is obtained by combining
Eqs. (\ref{mass1}) and (\ref{v}).
\begin{eqnarray}
\bar{m}_{\pi}^2&=& - \frac{N+2}{12} \lambda T^2 \nonumber \\
\bar{m}_{\sigma}^2&=& 2\lambda f_{\pi}^2 - \frac{N+2}{4} \lambda T^2
\label{mass2}
\end{eqnarray}
The full one loop self-energies for pions and the sigma meson are
drawn in Figures 1 and 2.  If one chooses $\langle \sigma \rangle = 0$
then there are no 1PR diagrams and the tadpoles should not be
included; they are already included in the temperature dependence
of $v$.  One may check this by fixing $v = f_{\pi}$ and then
computing the tadpole contributions to the effective masses.
One gets precisely Eq. (\ref{mass2}).  The diagrams involving
the 4-point vertices contribute an amount $(N+2) \lambda T^2/12$
to both the pion and sigma meson self-energies.  When evaluated
in the high temperature approximation and at low frequency and
momentum the 1PI diagrams involving the 3-point vertices may
be neglected.  (This follows from power counting.  These diagrams
involve two propagators instead of one, and so are only
logarithmically divergent in the UV in the vacuum.  The other
diagrams are quadratically divergent, which leads to a $T^2$
behavior at finite temperature.)
When all contributions of order $\lambda T^2$ are included
the pole positions of the pion and sigma
propagators move with the result that below $T_{\rm c}$
\begin{eqnarray}
m_{\pi}^2 &=& \bar{m}_{\pi}^2 + \Pi_{\pi} = 0 \nonumber \\
m_{\sigma}^2 &=& \bar{m}_{\sigma}^2 + \Pi_{\sigma} =
2 \lambda f_{\pi}^2 \left( 1 - T^2/T_{\rm c}^2 \right) \, ,
\end{eqnarray}
and above $T_{\rm c}$
\begin{equation}
m_{\pi}^2 = m_{\sigma}^2 = m_{\Phi}^2 = - \lambda f_{\pi}^2
+ \Pi_{\Phi} = \frac{N+2}{12} \lambda \left( T^2 - T_{\rm c}^2 \right) \, .
\end{equation}
The Goldstone Theorem is satisfied, there are no tachyons, and
restoration of the full symmetry of the Lagrangian above $T_{\rm c}$
is evident.

The second major problem is that long wavelength fluctuations
very near the phase transition cannot be treated with perturbation
theory because the self-interacting boson fields become massless
just at the transition.  Although this is a well--known problem
in the statistical mechanics of second order phase transitions, exactly
how it affects the critical temperature is not known for the linear
sigma model in 3 + 1 dimensions.  This is a topic for
further study.  The result presented here must be accepted for what it
is: a one loop estimate of the critical temperature.

\subsection{Nonlinear sigma model}

The nonlinear sigma model may be defined by the Lagrangian
\begin{equation}
{\cal L} = \frac{1}{2}\left(\partial_{\mu}{\bf \Phi}\right)^2
\end{equation}
together with the constraint
\begin{equation}
f_{\pi}^2 = {\bf \Phi}^2({\bf x},t) \, .
\end{equation}
The partition function is
\begin{equation}
Z = \int \left[ d{\bf \Phi}\right] \delta\left(f_{\pi}^2 - {\bf \Phi}^2
\right) \exp\left\{ \int_0^{\beta} d\tau \int_V d^3x \, {\cal L}\right\} \, .
\end{equation}
Because the length of the chiral field is fixed and cannot be changed
by thermal fluctuations it is often said that chiral symmetry breaking
is built into this model and therefore there can be no chiral symmetry
restoring phase transition.  On the other hand, the linear sigma model
does undergo a symmetry restoring phase transition.  Taking the quartic
coupling constant $\lambda$ to infinity essentially constrains the
length of the chiral field to be $f_{\pi}$ just as in the nonlinear
model.  The critical temperature, however, is independent of $\lambda$
at least in the mean field approximation.  So it would seem that the
phase transition survives.  If this is true, then one ought to be able
to derive it entirely within the context of the nonlinear model.
That is what we shall do, although it involves a lot more effort than
treatment of the linear model in the mean field approximation.
Since the only parameter in the model is $f_{\pi}$, and we are interested
in temperatures comparable to it, we cannot do an expansion in
powers of $T/f_{\pi}$.  The only other parameter is $N$, the number of
field components.  This suggests an expansion in $1/N$.

Begin by representing the field-constraining delta function by an
integral.
\begin{equation}
Z = \int \left[ d{\bf \Phi}\right] \left[ db'\right]
\exp\left\{ \int_0^{\beta} d\tau \int_V d^3x
\left[ {\cal L} +ib' \left( {\bf \Phi}^2 - f_{\pi}^2 \right)
\right]\right\}
\end{equation}
As with the linear model, we define the first
$N-1$ components of ${\bf \Phi}$ to be the pion field and the last
component to be the sigma field.  We allow for a zero
frequency and zero momentum condensate of the sigma field referred to
as $v$.  Following Polyakov \cite{Polyakov} we also separate out
explicitly the zero frequency and zero momentum mode of the
auxiliary field $b'$.  Integrating over all the other modes
will give us an effective action involving the constant part
of the fields.  We will then minimize the free energy with
respect to these constant parts, which is a saddle point
approximation.  Integrating over fluctuations about the saddle
point is a finite volume correction and of no consequence in the
thermodynamic limit.  The Fourier expansions are
\begin{eqnarray}
\Phi_i({\bf x},\tau)&=& \pi_i({\bf x},\tau) =
\sqrt{\frac{\beta}{V}} \sum_n \sum_{{\bf p}}
{\rm e}^{i({\bf x \cdot p}+\omega_n \tau)} \, {\tilde \pi}_i({\bf p},n)
\, ,\nonumber \\
\Phi_N({\bf x},\tau)&=& v + \sigma({\bf x},\tau) =
v + \sqrt{\frac{\beta}{V}} \sum_n \sum_{{\bf p}}
{\rm e}^{i({\bf x \cdot p}+\omega_n \tau)} \, {\tilde \sigma}({\bf p},n)
\, ,\nonumber \\
b'({\bf x},\tau)&=& i\frac{m^2}{2} + b({\bf x},\tau)
= i\frac{m^2}{2} + T \sqrt{\frac{\beta}{V}} \sum_n \sum_{{\bf p}}
{\rm e}^{i({\bf x \cdot p}+\nu_n \tau)} \, {\tilde b}({\bf p},n) \, .
\end{eqnarray}
One must remember to exclude the zero frequency and zero momentum
mode from the summations.
The field ${\bf \Phi}$ must be periodic in imaginary time for the usual
reasons, but there is no such requirement on the auxiliary
field $b$, hence $\omega_n = 2\pi n T$ and $\nu_n = \pi n T$.
Since the field $b$ has dimensions of inverse length squared we
inserted another factor of $T$ so as to make its Fourier amplitude
dimensionless, as they are for the other fields.  The action then
becomes
\begin{eqnarray}
S &=& \int_0^{\beta} d\tau \int_V d^3x \left\{ \frac{1}{2} \left[
\left(\partial_{\mu}\mbox{\boldmath $\pi$}\right)^2
- m^2 \mbox{\boldmath $\pi$}^2
+ \left(\partial_{\mu}\sigma\right)^2 -m^2 \sigma^2 \right]
- ib \left( 2v\sigma + \mbox{\boldmath $\pi$}^2 + \sigma^2 \right)
\right\} \nonumber \\
&+& \frac{1}{2}m^2 \left(f_{\pi}^2-v^2\right) \beta V
\end{eqnarray}
Note that terms linear in the fields integrate to zero
because $\langle \pi_i \rangle = \langle \sigma \rangle
= \langle b \rangle = 0$.

An effective action is derived by expanding exp($S$)
in powers of $b$ and integrating over the pion and sigma fields.
The term linear in $b$ vanishes on account of $\tilde{b}({\bf 0},0)
\propto \langle b \rangle = 0$.  The term proportional to $b^2$
is not zero and is exponentiated, thus summing a whole series
of contributions.  The term proportional to $b^3$ is not zero
either and it, too, may be exponentiated, summing an infinite series
of higher order terms left out of the order $b^2$ exponentiation.
After making the scaling $b \rightarrow b/\sqrt{2N}$ the effective
action becomes
\begin{eqnarray}
S_{\rm eff} &=&
- \frac{1}{2}\sum_n \sum_{\bf p}\,
\left( \omega_n^2 + p^2 + m^2 \right)
\left[ \tilde{ \mbox{\boldmath $\pi$}}({\bf p},n) \cdot
\tilde{ \mbox{\boldmath $\pi$}}(-{\bf p},-n) +
\tilde{ \sigma}({\bf p},n)\tilde{ \sigma}(-{\bf p},-n) \right]
\nonumber \\
&-&\frac{1}{2}\sum_n \sum_{\bf p}\, \left[ \Pi(p, \omega_n, T, m)
+\frac{2}{N} \, \frac{v^2}{\omega_n^2 + p^2 + m^2} \right]
\tilde{b}({\bf p},2n)\tilde{b}(-{\bf p},-2n)
\nonumber \\
&+& \frac{1}{2}m^2 \left(f_{\pi}^2-v^2\right) \beta V
+ {\rm O}\left( \tilde{b}^3/\sqrt{N}\right) \, .
\label{Seff}
\end{eqnarray}
Note that only even Matsubara frequencies contribute in the $b$--field:
$\nu_n = 2\pi nT$.  This may have been anticipated.
There appears the one loop function
\begin{equation}
\Pi(p, \omega_n, T, m) =T \sum_l \int \frac{d^3k}{(2\pi)^3} \,
\frac{1}{(\omega_n - \omega_l)^2 + ({\bf p}-{\bf k})^2 +m^2}
\, \frac{1}{\omega_l^2 + k^2 +m^2} \, .
\end{equation}
The effective action is an infinite series in $b$.  The coefficients
are frequency and momentum dependent, arising from one loop diagrams.
The coefficient of the term quadratic in $b$ in $S_{\rm eff}$
is illustrated schematically in Figure 3.
In addition, each successive term is suppressed by $1/\sqrt{N}$
compared to the previous one.  This is the large $N$ expansion.

The propagators for the pion and sigma fields are of the usual form
\begin{equation}
D_0^{-1}(p,\omega_n,m) = \omega_n^2 +p^2 + m^2
\end{equation}
with an effective mass $m$ yet to be determined.  The propagator
for the $b$--field is more complicated, being
\begin{equation}
D_b^{-1}(p, \omega_n,m) = \Pi(p, \omega_n, T, m)
+\frac{2}{N} \, \frac{v^2}{\omega_n^2 +p^2 + m^2} \, .
\end{equation}
The value of the condensate $v$ is not yet determined either.

Keeping only the terms up to order $b^2$ in $S_{\rm eff}$ (the rest
vanish in the limit $N \rightarrow \infty$) allows us to obtain an
explicit expression for the partition function and the pressure.
This includes the next to leading order in $N$.
\begin{eqnarray}
P &=& \frac{T}{V}\,\ln Z = \frac{1}{2}m^2 \left(f_{\pi}^2-v^2\right)
\nonumber \\
&-& \frac{N}{2}\,T \sum_n \int \frac{d^3p}{(2\pi)^3} \,
\ln\left[\beta^2 \left(\omega_n^2+p^2+m^2 \right) \right] \nonumber \\
&-& \frac{1}{2} T \sum_n \int \frac{d^3p}{(2\pi)^3} \,
\ln\left[ \Pi(p, \omega_n, T, m)
+ \frac{2}{N} \, \frac{v^2}{\omega_n^2 + p^2 + m^2}\right]
\end{eqnarray}
The second term under the last logarithm should and
will be set to zero at this order.  It may be needed at higher order
in the large $N$ expansion to regulate infrared divergences.

The pressure is extremized with respect to the mass parameter $m$.
Therefore $\partial P/\partial m^2 = 0$.  From the initial expression
for $Z$ this is seen to be equivalent to the thermal average of the
constraint.
\begin{equation}
f_{\pi}^2 = \langle {\bf \Phi}^2 \rangle =
v^2 + \langle \mbox{\boldmath $\pi$}^2 \rangle
+ \langle \sigma^2 \rangle
\end{equation}
If an approximation to the exact partition function is made, like
the large $N$ expansion, this constraint should still be satisfied.
It may, in fact, single out a preferred value of $m$.

To leading order in $N$ we may neglect the term involving $\Pi$
entirely.  The pressure is then
\begin{equation}
P = \frac{1}{2}m^2 \left(f_{\pi}^2-v^2\right) + N\,P_0(T,m) \, .
\end{equation}
The pressure must be a maximum with respect to variations in the
condensate $v$.  This means that
\begin{equation}
\partial P/\partial v = - m^2 v = 0 \, ,
\end{equation}
which is equivalent to the condition that $\langle \sigma \rangle = 0$.
There are two possibilities.
\begin{enumerate}
\item $m= 0$:  There exist massless particles, or Goldstone bosons,
and the value of the condensate is determined by the thermally
averaged constraint.  This is the symmetry--broken phase.
\item $v = 0$: The thermally averaged constraint is satisfied
by a nonzero temperature-dependent mass.  There are no Goldstone
bosons.  This is the symmetry--restored phase.
\end{enumerate}
Evidently there is a chiral symmetry restoring phase transition!

In the leading order of the large $N$ approximation the particles
are represented by free fields with a potentially temperature--dependent
mass $m$.  For any free bosonic field $\phi$
\begin{equation}
\partial P_0(T,m)/\partial m^2 = \langle \phi^2 \rangle =
\int \frac{d^3p}{(2\pi)^3} \, \frac{1}{\omega} \,
\frac{1}{{\rm e}^{\beta \omega} - 1}
\end{equation}
with $\omega = \sqrt{p^2 + m^2}$.  Thus extremizing the pressure with respect
to $m^2$ is equivalent to satisfying the thermally averaged constraint.
\begin{equation}
f_{\pi}^2 = v^2 + \langle \mbox{\boldmath $\pi$}^2 \rangle
+ \langle \sigma^2 \rangle
\end{equation}
Note however that the pion and sigma fields have the same mass and
therefore $\langle \mbox{\boldmath $\pi$}^2 \rangle =
(N-1) \langle \sigma^2 \rangle$.  Consider now the two different phases.

In the asymmetric phase the mass is zero.  The constraint is satisfied
by a temperature--dependent condensate.
\begin{equation}
v^2(T) = f_{\pi}^2 - \frac{N\, T^2}{12}
\end{equation}
This condensate goes to zero at a critical temperature
\begin{equation}
T_{\rm c}^2 = \frac{12}{N} f_{\pi}^2 \,\,\,\,\,
({\rm leading} \,\, N \,\, {\rm approximation}) \, .
\end{equation}
Exactly at $T_{\rm c}$ the thermally averaged constraint is satisfied
by the fluctuations of $N$ massless degrees of freedom without the help
of a condensate.

In the symmetric phase the condensate is zero.  The constraint is
satisfied by thermal fluctuations alone.
\begin{equation}
f_{\pi}^2 = N \int \frac{d^3p}{(2\pi)^3} \, \frac{1}{\omega} \,
\frac{1}{{\rm e}^{\beta \omega} - 1}
\end{equation}
Thermal fluctuations decrease with increasing mass at fixed temperature.
The constraint is only satisfied by massless excitations at one
temperature, namely, $T_{\rm c}$.  At temperatures $T > T_{\rm c}$
the mass must be greater than zero.  Near the critical temperature
the mass should be small, and the flucutations may be expanded
about $m = 0$ as
\begin{equation}
f_{\pi}^2 = N T^2 \left[ \frac{1}{12} - \frac{m}{4\pi T}
-\frac{m^2}{8\pi^2 T^2} \ln\left( \frac{m}{4\pi T} \right)
- \frac{m^2}{16\pi^2 T^2} + \cdots \right] \, .
\end{equation}
As $T$ approaches $T_{\rm c}$ from above, the mass approaches zero like
\begin{equation}
m(T) = \frac{\pi}{3T} \left(T^2 - T_{\rm c}^2 \right) + \cdots \, .
\end{equation}
This is a second order phase transition since there is no possibility
of metastable supercooled or superheated states.

The mass must grow faster than the temperature at very high temperatures
in order to keep the field fluctuations fixed and equal to $f_{\pi}^2$.
Asymptotically the particles move nonrelativistically.
This allows us to compute the fluctuations analytically.  We get
\begin{equation}
f_{\pi}^2 = N \left( \frac{T}{2\pi} \right)^{3/2} \sqrt{m}
\,\, {\rm e}^{-m/T} \, .
\end{equation}
This is a transcendental equation for $m(T)$.  It can also be written as
\begin{equation}
m = T \, \ln \left( \frac{NT}{2\pi f_{\pi}} \sqrt{\frac{mT}{2\pi f_{\pi}^2}}
\right) \, .
\end{equation}
Roughly, the solution behaves as
\begin{equation}
m \sim T \, \ln \left( T^2/T_{\rm c}^2 \right) \, .
\end{equation}

It is rather amusing that, in leading order of the large $N$ approximation,
the elementary excitations are massless below $T_{\rm c}$, become
massive above $T_{\rm c}$, and at asymptotically high temperatures
move nonrelativistically.

The result to first order of the large $N$ expansion provides good
insight into the nature of the two--phase structure of the nonlinear
sigma model, but it is not quite satisfactory for two reasons.
First, it predicts $N$ massless Goldstone bosons in the broken
symmetry phase when in fact we know there ought to be only $N - 1$.
Second, the square of the critical temperature is $12 f_{\pi}^2/N$
whereas it is $12 f_{\pi}^2/(N+2)$ in the linear sigma model in
mean field approximation; we expect them to be the same in
the limit $\lambda \rightarrow \infty$.  Both these problems can
be rectified by inclusion of the next to leading order term in
$N$, namely, the contribution of the $b$--field.

It is natural to expect that the $b$--field will contribute essentially
one negative degree of freedom to the $T^4$ term in the pressure so
as to give $N-1$ Goldstone bosons in the low temperature phase.
Therefore we move one of the $N$ degrees of freedom and put it
together with the $b$ contribution as
\begin{eqnarray}
P &=& \frac{1}{2}m^2 \left(f_{\pi}^2-v^2\right)
- \frac{N-1}{2}\,T \sum_n \int \frac{d^3p}{(2\pi)^3} \,
\ln\left[\beta^2 \left(\omega_n^2+p^2+m^2 \right) \right] \nonumber \\
&-& \frac{1}{2} T \sum_n \int \frac{d^3p}{(2\pi)^3} \,
\ln\left[ \beta^2 \left(\omega_n^2+p^2+m^2 \right) \Pi \right] \, .
\end{eqnarray}
The function $\Pi(p,\omega_n,T,m)$ can be reduced to a
one dimensional integral
\begin{equation}
\Pi = \frac{1}{8\pi^2p} \int_0^{\infty} \frac{dk\,k}{\omega}
\ln\left[ \frac{k^2+pk+\Lambda^2}{k^2-pk+\Lambda^2} \right]
\,\frac{1}{{\rm e}^{\beta \omega}-1}
\end{equation}
where
\begin{equation}
\Lambda^2 = \Lambda^2(p,\omega_n,m) =
\frac{(\omega_n^2+p^2)^2 + 4m^2\omega_n^2}{4(\omega_n^2+p^2)}
\end{equation}
but unfortunately cannot be simplified any further.
In any case, to the order in $N$ to which we are working, the pressure
is
\begin{equation}
P = \frac{1}{2}m^2 \left(f_{\pi}^2-v^2\right)
+ (N-1)\,P_0(T,m) + P_{\rm I}(T,m) \, .
\end{equation}
This can be thought of, in the low temperature phase, as $N - 1$
Goldstone bosons with an interaction term $P_{\rm I}$.

Because of the logarithm the main contribution to the interaction
pressure will come when $\Pi$ is very small compared to one.  This
corresponds to very large values of the parameter $\Lambda$;
in other words, to very high momentum, Matsubara frequency, or mass.
In this limit
\begin{equation}
\Pi \rightarrow \frac{1}{4\pi^2\Lambda^2}\,
\int_0^{\infty} \frac{dk\,k^2}{\omega}
\,\frac{1}{{\rm e}^{\beta \omega}-1}
= \frac{h_3(m/T)}{4\pi^2}\,\frac{T^2}{\Lambda^2} \, .
\end{equation}
This may be thought of as a kind of high energy approximation,
and we shall henceforth refer to it as such.  Then
\begin{eqnarray}
P_{\rm I} &=& \frac{1}{2}\,T \sum_n \int \frac{d^3p}{(2\pi)^3} \,
\ln\left[\beta^2 \left(\omega_n^2+p^2+m^2 \right) \Pi \right] \nonumber \\
&\approx& - \frac{1}{2}\,T \sum_n \int \frac{d^3p}{(2\pi)^3} \,
\ln\left[\frac{h_3}{\pi^2}\, \frac{(\omega_n^2+p^2)(\omega_n^2+p^2+m^2)}
{(\omega_n^2+\omega_+^2)(\omega_n^2+\omega_-^2)} \right]
\end{eqnarray}
with the dispersion relations
\begin{equation}
\omega_{\pm}^2 = p^2 +2m^2 \pm 2m\sqrt{p^2+m^2} \, .
\end{equation}
The interaction pressure can now be determined in the usual way to be
\begin{displaymath}
P_{\rm I} = - T \int \frac{d^3p}{(2\pi)^3} \left\{
\ln\left[ 1- {\rm e}^{-\beta p} \right]
+ \ln\left[ 1- {\rm e}^{-\beta \omega(p)} \right] \right.
\end{displaymath}
\begin{equation}
\left. - \ln\left[ 1- {\rm e}^{-\beta \omega_+(p)} \right]
- \ln\left[ 1- {\rm e}^{-\beta \omega_-(p)} \right] \right\} \, .
\end{equation}
Note that $h_3(m/T)$ has no effect within this approximation.
Note also that in the broken symmetry phase where $m = 0$ the contribution
of the $b$--field cancels one of the massless degrees of freedom
to give $N-1$ Goldstone bosons.

Now we are prepared to examine the behavior of the system near
the critical temperature with the inclusion of next-to-leading
terms in $N$.  We do an expansion in $m/T$ as before.  The pressure
is, up to and including order $m^3$:
\begin{equation}
P = (N-1)\,\frac{\pi^2}{90}T^4 - \frac{N+2}{24} m^2T^2
+\frac{1}{2}m^2 \left(f_{\pi}^2-v^2\right)
+ \frac{N}{12 \pi} m^3 T \, .
\end{equation}
In the high temperature phase where $v = 0$ maximization with
respect to $m$ yields
\begin{equation}
f_{\pi}^2 = T^2 \left[ \frac{N+2}{12} - \frac{N}{4\pi} \frac{m}{T}
\right] \, .
\end{equation}
This gives the same critical temperature as in the mean field
treatment of the linear sigma model.
\begin{equation}
T_{\rm c}^2 = \frac{12}{N+2}\,f_{\pi}^2  \,\,\,\,\,
({\rm subleading} \,\, N \,\, {\rm approximation}) \, .
\end{equation}
The mass approaches zero from above like
\begin{equation}
m(T) = \frac{\pi (N+2)}{3 N T} \left( T^2- T_{\rm c}^2 \right) \, .
\end{equation}
We leave it as an exercise for the reader to compute the asymptotic
behavior of the mass with the inclusion of the subleading terms in
$N$.

The results obtained immediately above used an approximation for
$\Pi$ which we referred to as a high energy approximation.
Relaxing this approximation can be done albeit at the cost of
a numerical calculation.  We do not attempt that in this paper.
Of course, one should also go beyond the mean field approximation
in the linear model.

\section{$f_{\pi}$ at Low Temperature}

Consideration of correlation functions at finite temperature
is more involved than at zero temperature.
Lorentz invariance is not manifest because there
is a preferred frame of reference, the frame in which the matter is
at rest.  Thus spectral densities and other functions may depend on
energy and momentum separately and not just on their invariant $s$.
Also, the number of Lorentz tensors is greater because there is a new
vector available, namely, the vector $u_{\mu}$ = (1,0,0,0) which
specifies the rest frame of the matter.

For a given four-momentum $q$ it is useful to define two projection tensors.
The first one, $P_T^{\mu\nu}$, is both three- and four-dimensionally
transverse,
\begin{equation}
P_T^{ij} \, \equiv \, \delta^{ij} \, - \, \frac{q^iq^j}{{\bf q}^2} \, ,
\end{equation}
with all other components zero.  The second one, $P_L^{\mu\nu}$,
is only four-dimensionally transverse,
\begin{equation}
P_L^{\mu\nu} \, \equiv \, -\left( g^{\mu\nu} \, - \, \frac{q^{\mu}q^{\nu}}
{q^2} \, + \, P_T^{\mu\nu} \right) \, .
\end{equation}
The notation is $L$ for longitudinal and $T$ for transverse with respect
to ${\bf q}$. There are no other symmetric second rank tensors which
are four-dimensionally transverse.

In the usual fashion \cite{Frad,Wal} one may construct a Green function
for the axial vector current
\begin{equation}
G_{ab}^{\mu\nu}(z,{\bf q}) \,=\, \int_{-\infty}^{\infty}
\frac{d\omega}{\omega - z} \; \rho^{\mu\nu}_{ab} (\omega,{\bf q}) \, ,
\end{equation}
where the spectral density tensor is
\begin{eqnarray}
\rho^{\mu\nu}_{ab} (\omega,{\bf q}) &=& \frac{1}{Z}
\sum_{m,n} (2\pi)^3 \delta(\omega -E_m +E_n)
\delta({\bf q} -{\bf p_m} +{\bf p_n}) \nonumber \\
& \times & \left({\rm e}^{-E_n/T} - {\rm e}^{-E_m/T} \right)
\langle n| {\cal A}^{\mu}_a (0) |m \rangle
\langle m| {\cal A}^{\nu}_b (0) |n \rangle \, .
\end{eqnarray}
The summation is over a complete set of energy eigenstates.
The retarded, advanced and Matsubara Green functions are
\begin{eqnarray}
G_{ab}^{{\rm R}\, \mu\nu}(q_0,{\bf q})
&=& G_{ab}^{\mu\nu}(q_0 + i\epsilon,{\bf q}) \, ,\\
G_{ab}^{{\rm A}\, \mu\nu}(q_0,{\bf q})
&=& G_{ab}^{\mu\nu}(q_0 - i\epsilon,{\bf q}) \, ,\\
G_{ab}^{{\rm T}\, \mu\nu}(\omega_n,{\bf q})
&=& G_{ab}^{\mu\nu}(i\omega_n,{\bf q}) \, ,
\end{eqnarray}
where $\epsilon \rightarrow 0^+$.

Due to current conservation the spectral density tensor can
be decomposed into longitudinal and transverse pieces \cite{Wsum}.
\begin{equation}
\rho^{\mu\nu}_{ab}(q) \,=\, \delta_{ab} \left[ \rho_A^L(q)
P_L^{\mu\nu} + \rho_A^T(q) P_T^{\mu\nu} \right]
\end{equation}
In general the spectral densities depend on $q^0$ and ${\bf q}$ separately
as well as on the temperature.
In the vacuum we can always go to the rest frame of a massive particle,
and in that frame there can be no difference between longitudinal and
transverse polarizations, so that $\rho_L = \rho_T = \rho$.  We also
observe that $P_L^{\mu\nu}+P_T^{\mu\nu} = -(g^{\mu\nu} - q^{\mu}
q^{\nu}/q^2)$.  The pion, being a massless
Goldstone boson, is special.  It contributes to the longitudinal axial
spectral density and not to the transverse one.  In vacuum
\begin{equation}
\rho^{\mu\nu}(q) \, = \,
\left( \frac{q^{\mu}q^{\nu}}{q^2} - g^{\mu\nu} \right)
\rho_A(q^2) \, + \, f_{\pi}^2 \delta(q^2) q^{\mu}q^{\nu} \, .
\end{equation}
This may be taken to be the definition of the pion decay constant
at zero temperature.  In fact, one can write the pion's contribution as
\begin{equation}
f_{\pi}^2 \delta(q^2) q^{\mu}q^{\nu} \, = \, f_{\pi}^2 q^2 \delta(q^2)
P_L^{\mu\nu} \, .
\end{equation}
This cannot be taken as the definition of the pion decay constant
at {\em finite temperature} because the contribution
of the pion to the longitudinal spectral density cannot be assumed
to be a $\delta$ function in $q^2$.  In general the pion's dispersion
relation will be more complicated and will develop a width at nonzero
momentum.  This smears out the delta function into something like
a relativistic Breit-Wigner distribution.
Fortunately, the Goldstone Theorem \cite{Gold} requires that there
be a zero frequency excitation when the momentum is zero.
(For a proof applicable to relativistic quantum field theories at
finite temperature see \cite{JK}.)
This implies that the width must go to zero at ${\bf q} = {\bf 0}$
which results in a delta function at zero frequency.
Explicit calculations support this assertion
\cite{Goity1,Schenk,Song}.
Therefore it seems to make sense to define
\begin{equation}
f_{\pi}^2(T) \equiv  2 \lim_{\epsilon \rightarrow 0}
\int_0^{\epsilon} \frac{d q_0^2}{q_0^2}
\, \rho_A^L(q_0, {\bf q} = {\bf 0}) \, .
\end{equation}
Physically this means that the pion decay constant at finite
temperature measures the strength of the coupling of the Goldstone
boson to the longitudinal part of the retarded axial vector response
function in the limit of zero momentum.

We shall study the pion's contribution to the spectral density
only at temperatures small compared to $f_{\pi}$.
We shall study both the nonlinear and the
linear sigma models.  At low temperatures the sigma meson's
contribution as a material degree of freedom is frozen out
and one might expect the same dynamics to be operative in both
models; in other words, one may expect the result to be the same
and so independent of $\lambda$.  For temperatures approaching
$T_{\rm c}$ the problem is more difficult and is
left for future investigation.

\subsection{Nonlinear sigma model}

The nonlinear sigma model was defined at the beginning of section
2.2.  One can make a nonlinear redefinition of the field without
changing the physical content of the theory.  Various redefinitions
may be found in the literature.  We will first list the most common
ones, and then we will compute $f_{\pi}(T)$ for each of them, thereby
illustrating that one always gets the same result.  It is interesting
to see how this comes about; it is also reassuring that it does.

A convenient way to express the sigma and pion fields which
explicitly contains the constraint is
\begin{eqnarray}
\sigma &=& f_{\pi} \cos\left(\phi/f_{\pi}\right) \nonumber \\
\mbox{\boldmath $\pi$} &=& f_{\pi} \hat{\mbox{\boldmath $\phi$}}
\sin\left(\phi/f_{\pi}\right) \, ,
\end{eqnarray}
where $\phi = |\mbox{\boldmath $\phi$}|$ and
$\hat{\mbox{\boldmath $\phi$}} = \mbox{\boldmath $\phi$} /\phi$.
The Lagrangian may then be expressed in terms of the fields of choice.
\begin{eqnarray}
{\cal L} &=& \frac{1}{2} \partial_{\mu} \mbox{\boldmath $\pi$}
\cdot \partial^{\mu} \mbox{\boldmath $\pi$}
+ \frac{1}{2} \partial_{\mu} \sigma \, \partial^{\mu} \sigma \nonumber \\
&=& \frac{1}{2} \partial_{\mu} \mbox{\boldmath $\pi$}
\cdot \partial^{\mu} \mbox{\boldmath $\pi$}
+ \frac{1}{2} \frac{\left( \mbox{\boldmath $\pi$}\cdot
\partial_{\mu} \mbox{\boldmath $\pi$} \right)
\left(\mbox{\boldmath $\pi$}\cdot
\partial^{\mu} \mbox{\boldmath $\pi$} \right)}
{f_{\pi}^2 - \pi^2} \nonumber \\
&=& \frac{1}{2} \, \frac{f_{\pi}^2}{\phi^2} \,
\sin^2\left( \frac{\phi}{f_{\pi}} \right)
\partial_{\mu} \mbox{\boldmath $\phi$}
\cdot \partial^{\mu} \mbox{\boldmath $\phi$}
+ \frac{1}{2} \left[ 1 - \frac{f_{\pi}^2}{\phi^2}
\sin^2\left( \frac{\phi}{f_{\pi}} \right) \right]
\partial_{\mu} \phi \, \partial^{\mu} \phi
\end{eqnarray}
Another representation to consider is due to Weinberg \cite{W1},
who defines
\begin{equation}
{\bf p} = 2\, \frac{f_{\pi}^2}{\pi^2} \,
\left( 1 - \sqrt{1 - \frac{\pi^2}{f_{\pi}^2}} \right)
\mbox{\boldmath $\pi$}
\end{equation}
or inversely
\begin{equation}
\mbox{\boldmath $\pi$} = \frac{{\bf p}}{1 + p^2/4f_{\pi}^2} \, .
\end{equation}
In terms of Weinberg's field definition the Lagrangian is very compact:
\begin{equation}
{\cal L} = \frac{1}{2} \, \frac{ \partial_{\mu} {\bf p}
\cdot \partial^{\mu} {\bf p}}{\left( 1 + p^2/4f_{\pi}^2 \right)^2}
\, .
\end{equation}

The $(\sigma, \mbox{\boldmath $\pi$})$ representation is cumbersome
because of the constraint, although it can be handled by the Lagrange
multiplier method of section 2.  However, it is inconvenient
for exposing the physical particle content and for doing perturbation
theory in terms of physical particles.  Among the three physical
representations we choose to work with here, it is interesting to
note the range of allowed values of the fields.  The magnitude of
the ${\bf p}$--field can range from zero to infinity, the
magnitude of the $\mbox{\boldmath $\pi$}$--field
can range from 0 to $f_{\pi}$, and the magnitude of the
$\mbox{\boldmath $\phi$}$--field can range from 0
to $\pi f_{\pi}$.  This distinction is important when dealing
with nonperturbative large amplitude motion; whether it makes
any difference in low orders of perturbation theory is not
known to us.

The first step in our quest to extract the temperature dependence
of $f_{\pi}$ from the theory is to obtain the form of the
axial vector current in terms of the chosen fields.  Starting
from
\begin{equation}
{\bf {\cal A}}_{\mu} =  - \sigma \, \partial_{\mu} \mbox{\boldmath $\pi$}
+ \mbox{\boldmath $\pi$}\, \partial_{\mu} \sigma
\end{equation}
one directly computes
\begin{eqnarray}
{\bf {\cal A}}_{\mu} &=& - \sigma \left[ \partial_{\mu} \mbox{\boldmath $\pi$}
+ \frac{ \mbox{\boldmath $\pi$} \,
\left(\mbox{\boldmath $\pi$}\cdot
\partial_{\mu} \mbox{\boldmath $\pi$} \right)}
{f_{\pi}^2 - \pi^2} \right] \nonumber \\
&=& - \frac{f_{\pi}^2}{2 \phi} \,
\sin\left( \frac{2 \phi}{f_{\pi}} \right)
\partial_{\mu} \mbox{\boldmath $\phi$}
- f_{\pi} \hat{\mbox{\boldmath $\phi$}}
\left[ 1 - \frac{f_{\pi}}{2 \phi}
\sin\left( \frac{2 \phi}{f_{\pi}} \right) \right]
\hat{\mbox{\boldmath $\phi$}}
\cdot \partial_{\mu} \mbox{\boldmath $\phi$} \nonumber \\
&=& - \frac{1}{f_{\pi}} \, \frac{1}{(1+p^2/4f_{\pi}^2)^2}
\left[ \left( f_{\pi}^2 - \frac{1}{4}p^2 \right) \partial_{\mu}
{\bf p} + \frac{1}{2} {\bf p} \left( {\bf p} \cdot \partial_{\mu}
{\bf p} \right) \right] \, .
\end{eqnarray}
Every form of the axial vector current is an odd function of
the pion field.

Obviously it is not possible to
compute the axial vector correlation function exactly.  We will
restrict our attention to low temperature.  Roughly speaking,
a loop expansion of the correlation function is an expansion
in powers of $T^2/f_{\pi}^2$, with each additional loop contributing
one more such factor.  To one loop order we need the axial vector
current to third order in the pion field.
\begin{eqnarray}
{\bf {\cal A}}_{\mu} &=& - f_{\pi} \, \partial_{\mu}
\mbox{\boldmath $\pi$}
+ \frac{\pi^2}{2 f_{\pi}} \, \partial_{\mu} \mbox{\boldmath $\pi$}
- \frac{1}{f_{\pi}} \, \mbox{\boldmath $\pi$}
\left(\mbox{\boldmath $\pi$}\cdot
\partial_{\mu} \mbox{\boldmath $\pi$} \right) \nonumber \\
&=& - f_{\pi} \, \partial_{\mu} \mbox{\boldmath $\phi$}
+ \frac{2 \phi^2}{3 f_{\pi}} \, \partial_{\mu} \mbox{\boldmath $\phi$}
- \frac{2}{3f_{\pi}} \, \mbox{\boldmath $\phi$}
\left(\mbox{\boldmath $\phi$}\cdot
\partial_{\mu} \mbox{\boldmath $\phi$} \right) \nonumber \\
&=& - f_{\pi} \, \partial_{\mu} {\bf p}
+ \frac{3 p^2}{4 f_{\pi}} \, \partial_{\mu} {\bf p}
- \frac{1}{2f_{\pi}} \, {\bf p} \left( {\bf p} \cdot \partial_{\mu}
{\bf p} \right)
\end{eqnarray}
We will also need the Lagrangian to fourth order in the pion field.
\begin{eqnarray}
{\cal L}_4 &=& \frac{1}{2 f_{\pi}^2}
\left( \mbox{\boldmath $\pi$}\cdot
\partial_{\mu} \mbox{\boldmath $\pi$} \right)
\left(\mbox{\boldmath $\pi$}\cdot
\partial^{\mu} \mbox{\boldmath $\pi$} \right) \nonumber \\
&=& \frac{1}{6 f_{\pi}^2}\left[
\left( \mbox{\boldmath $\phi$}\cdot
\partial_{\mu} \mbox{\boldmath $\phi$} \right)
\left(\mbox{\boldmath $\phi$}\cdot
\partial^{\mu} \mbox{\boldmath $\phi$} \right)
- \phi^2 \, \partial_{\mu} \mbox{\boldmath $\phi$} \cdot
\partial^{\mu} \mbox{\boldmath $\phi$} \right] \nonumber \\
&=& - \frac{1}{4 f_{\pi}^2} \, p^2 \,
\partial_{\mu} {\bf p} \cdot \partial^{\mu} {\bf p}
\end{eqnarray}
The correlation function $\langle {\cal A}_{\mu}^i(x)\,{\cal A}
_{\nu}^j(y) \rangle$
will have a zero loop contribution from the pi--pi correlation
function $\langle \partial_{\mu}\pi^i(x)\,\partial_{\nu}\pi^j(y)
\rangle$, a one loop self--energy correction to the same pi--pi
correlation function, and a one loop contribution from the
correlation function $\langle \partial_{\mu}\pi^i(x)\,
\pi^j(y) \pi^k(y) \partial_{\nu}\pi^l(y) \rangle$ involving four pions.
These three contributions are illustrated in Figure 4.

The contribution of the bare pion propagator $D_0$ to the longitudinal
spectral density is easily found to be
\begin{equation}
\rho_A^L(q_0, {\bf q}) =
f_{\pi}^2 \, q^2 \, \delta \left( q^2 \right) \, .
\end{equation}
At zero temperature this is just the definition of the pion decay
constant.

The one loop pion self-energy may be computed by standard diagrammatic
or functional integral techniques.  The results are:
\begin{eqnarray}
\Pi_{\mbox{\boldmath $\pi$}}(q) &=& - \frac{T^2}
{12 f_{\pi}^2}\, q^2 \nonumber \\
\Pi_{\bf p}(q) &=& (N - 1) \, \frac{T^2}
{24 f_{\pi}^2}\, q^2 \nonumber \\
\Pi_{\mbox{\boldmath $\phi$}}(q) &=& \frac{1}{3} \,
\Pi_{\mbox{\boldmath $\pi$}}(q) + \frac{2}{3} \,
\Pi_{\bf p}(q) \, .
\label{self}
\end{eqnarray}
These are quite dependent on the definition of the pion field!
Nevertheless, it is worth noting that the Goldstone Theorem is satisfied
on account of the fact that the self-energy is always proportional
to $q^2$.

The final contribution comes from the correlation function of a pion
at point $x$ with three pions at point $y$.  Again, standard
diagrammatic or functional integral techniques may be used.
To express the answers, we gather together the contributions from
the bare propagator, from the one loop self-energy, and from
this correlation function, and quote the coefficient of the
term $f_{\pi}^2 \, q^2 \, \delta \left( q^2 \right)$ in the
longitudinal part of the axial vector spectral density.
\begin{eqnarray}
\mbox{\boldmath $\pi$}: && \left[ 1 - \frac{T^2}{12 f_{\pi}^2}
\right] - (N - 3) \frac{T^2}{12 f_{\pi}^2} \nonumber \\
{\bf p}: && \left[ 1 + (N-1) \frac{T^2}{24 f_{\pi}^2}
\right] - \left( N - \frac{5}{3} \right) \frac{T^2}{8 f_{\pi}^2}
\nonumber \\
\mbox{\boldmath $\phi$}: && \left[ 1 + (N - 2) \frac{T^2}{36 f_{\pi}^2}
\right] - (N - 2) \frac{T^2}{9 f_{\pi}^2} \nonumber \\
\end{eqnarray}
In all three cases the results are the same and amount to a
temperature dependence of
\begin{equation}
f_{\pi}^2(T) = f_{\pi}^2 \left[ 1 - \frac{N-2}{12} \, \frac{T^2}
{f_{\pi}^2} \right] \, .
\end{equation}
It agrees with Eq. (\ref{Gasfpi}) for the only case which they can
be compared: $N_f^2 = N = 4$.  The calculation of Gasser and
Leutwyler was verified by Eletsky and Kogan \cite{Kogan}.

\subsection{Linear sigma model}

It is now not surprising to discover that the linear sigma model
gives the same result for $f_{\pi}(T)$ at low temperature as the
nonlinear sigma model.  This is because the sigma meson is very
heavy at low temperature and cannot contribute materially the
way the pions do.  However, the way in which it works out is
very different.

Let us go back to the axial vector current {\em before} shifting
the sigma field.
\begin{equation}
{\bf {\cal A}}_{\mu} =  - \sigma \, \partial_{\mu} \mbox{\boldmath $\pi$}
+ \mbox{\boldmath $\pi$}\, \partial_{\mu} \sigma
\end{equation}
{\em After} making the shift $\sigma \rightarrow v + \sigma$ the
current takes the form
\begin{equation}
{\bf {\cal A}}_{\mu} = -v \, \partial_{\mu} \mbox{\boldmath $\pi$}
 - \sigma \, \partial_{\mu} \mbox{\boldmath $\pi$}
+ \mbox{\boldmath $\pi$}\, \partial_{\mu} \sigma \, .
\end{equation}
By maximizing the pressure (minimizing the effective potential)
with respect to $v$ at each temperature we effectively sum all
tadpole diagrams, leaving only 1PI diagrams in any subsequent
perturbative treatment.  If this is done, one's inclination is
to identify $v(T)$ with $f_{\pi}(T)$.  This is wrong; $f_{\pi}(T)$
has additional contributions, as we shall now see.

The first contribution to $f_{\pi}^2(T)$ does come from $v^2(T)$
since it involves the cross term of $\partial_{\mu} \pi^a(x)$
with $\partial_{\nu} \pi^a(y)$.  Following the analysis of
section 2.1, but at low temperature rather than high, we simply
leave out the contribution of the heavy sigma meson.  This gives
\begin{equation}
P(T,v) = (N-1) \frac{\pi^2}{90}T^4 + \frac{\lambda}{2}v^2 \left[
f_{\pi}^2 - \frac{N-1}{12}T^2\right] - \frac{\lambda}{4}v^4 \, .
\end{equation}
Maximizing with respect to $v$ gives
\begin{equation}
v^2(T) = f_{\pi}^2 - \frac{N-1}{12} T^2 \, .
\end{equation}
The $T^2/f_{\pi}^2$ correction is identically the tadpole contribution
to the vertex shown in Figure 5.

There is another, nonlocal, contribution to the vertex shown in
Figure 5, corresponding to the emission and absorption of a virtual
sigma meson.  One might think that it is suppressed by the large
sigma mass, $m_{\sigma}^2 = 2 \lambda f_{\pi}^2$, but in fact this
is compensated by the coupling constant $\lambda$ in the
extra vertex.  Evaluation of this diagram gives a contribution to
$f_{\pi}^2(T)$ of $T^2/6 f_{\pi}^2$.

Finally there is a contribution coming from the dressed pion
propagator analogous to the nonlinear sigma model.  The full
one loop 1PI pion self-energy diagrams were already shown in
Figure 1.  We know that the sum of the momentum independent
pieces is zero on account of Goldstone's Theorem.  We only need
the contribution which is quadratic in the energy and momentum of
the pion.  This can only arise from the so-called exchange
diagram involving two $\sigma \pi \pi$ vertices, also shown
in Figure 5.  In imaginary time (Euclidean space) it is
\begin{equation}
\Pi_{\rm ex}(\omega_n, {\bf q}) =
-4 \lambda^2 f_{\pi}^2 \, T \sum_l \int \frac{d^3k}{(2\pi)^3}
\, \frac{1}{\omega_l^2 + k^2} \, \frac{1}{(\omega_l + \omega_n)^2
+ ({\bf k} + {\bf q})^2 + m_{\sigma}^2} \, .
\end{equation}
Since $T \ll m_{\sigma}$ it is easy to extract the piece quadratic
in the momentum.  Analytically continuing to Minkowski space
$(\omega_n \rightarrow i q_0)$ it is $q^2 T^2/12 f_{\pi}^2$.

The residue of the pion pole in the axial vector correlation function
can now be obtained by adding the vacuum contribution, the pion
self-energy correction, and the tadpole and nonlocal vertex
corrections as follows.
\begin{displaymath}
\left[ 1  - \frac{1}{12}\frac{T^2}{f_{\pi}^2} \right]
- \frac{N-1}{12}\frac{T^2}{f_{\pi}^2}
+ \frac{1}{6}\frac{T^2}{f_{\pi}^2}
\end{displaymath}
The final result,
\begin{equation}
f_{\pi}^2(T) = f_{\pi}^2 \left[ 1 - \frac{N-2}{12} \, \frac{T^2}
{f_{\pi}^2} \right] \, ,
\end{equation}
is identical to that of the nonlinear sigma model.
We remark that this cannot be used to compute the critical
temperature since it was obtained under the condition that
$T \ll f_{\pi}$.

\section{Scalar Condensate at Low Temperature}

The scalar condensate is defined as $|\langle {\bf \Phi} \rangle |$.
Our convention has been to allow the last, $N$'th component of the
field to condense, and to refer to this as either $v$ (if the field
is shifted) or $\langle \sigma \rangle$ (if the field is not shifted).
In this section we use the latter convention.

It is interesting to ask what happens to this condensate as a function
of temperature in the nonlinear model.  The constraint as an operator
equation is $f_{\pi}^2 = {\bf \Phi}^2$ and as a thermal average is
$f_{\pi}^2 = \langle {\bf \Phi}^2 \rangle$; it is not
$f_{\pi} = |\langle {\bf \Phi} \rangle |$.  The condensate indeed
can change with temperature.  In fact we can quite easily compute it
to two loop order.  Before doing so, we first discuss the connection
of this condensate with the quark condensate $\langle \bar{q}q \rangle$.

In two flavor QCD one oftentimes associates the sigma and
pion fields with certain bilinears of the quark fields.
\begin{displaymath}
\bar{q}q \, \sim \, \sigma
\end{displaymath}
\begin{displaymath}
i \bar{q} \gamma_5 \mbox{\boldmath $\tau$} q \, \sim \,
\mbox{\boldmath $\pi$}
\end{displaymath}
This association is made because the quark bilinears transform in
the same way under SU(2)$\times$SU(2) as the corresponding meson
fields.  The dimensions don't match so there must be some
dimensionful coefficient relating them; this coefficient could even
be a function of the group invariant $\sigma^2 +
\mbox{\boldmath $\pi$}^2 \sim \left( \bar{q}q \right)^2
-\left( \bar{q} \gamma_5 \mbox{\boldmath $\tau$} q \right)^2$.
Does this particular combination of four--quark condensates change with
temperature?  The temperature dependence of the four--quark condensates
at low temperatures was first calculated in \cite{BSh} with the help of
the fluctuation-dissipation theorem. The contribution of pions alone was
later discussed in \cite{Eletsky3} using soft pion techniques.
>From \cite{BSh,Eletsky3} one can read off the two condensates
separately:
\begin{equation}
\langle \left( \bar{q}q \right)^2 \rangle =
\left[1 - \frac{T^2}{4f_{\pi}^2} \right]
\langle 0| \left( \bar{q}q \right)^2 |0 \rangle
- \frac{T^2}{12f_{\pi}^2} \langle 0|
\left( \bar{q} \gamma_5 \mbox{\boldmath $\tau$} q \right)^2
|0 \rangle
\end{equation}
and
\begin{equation}
\langle \left( \bar{q} \gamma_5 \mbox{\boldmath $\tau$} q \right)^2
\rangle = \left[1 - \frac{T^2}{12f_{\pi}^2} \right]
\langle 0| \left( \bar{q} \gamma_5 \mbox{\boldmath $\tau$} q \right)^2
|0 \rangle - \frac{T^2}{4f_{\pi}^2} \langle 0|
\left( \bar{q} q \right)^2
|0 \rangle \, .
\end{equation}
Therefore there is no correction to this group invariant to order
$T^2/f_{\pi}^2$ inclusive.
\begin{equation}
\langle \left( \bar{q} q \right)^2 -
\left( \bar{q} \gamma_5 \mbox{\boldmath $\tau$} q \right)^2 \rangle
\, = \, \langle 0| \left( \bar{q} q \right)^2 -
\left( \bar{q} \gamma_5 \mbox{\boldmath $\tau$} q \right)^2
|0 \rangle
\end{equation}
This result is consistent with our analysis of the nonlinear sigma
model in sections 2.2 and 3.1.

Now let us return to the business of computing the temperature
dependence of the scalar condensate to one and two loop order.  In
terms of the three representations used in section 3.1 the sigma
field is
\begin{eqnarray}
\sigma/f_{\pi} &=&
\sqrt{1 - \frac{\mbox{\boldmath $\pi$}^2}
{f_{\pi}^2}} = 1 - \frac{\mbox{\boldmath $\pi$}^2}
{2f_{\pi}^2} - \frac{\left(\mbox{\boldmath $\pi$}^2\right)^2}
{8f_{\pi}^4} + \cdots \nonumber \\
&=& \left[ 1 - \frac{{\bf p}^2}{2f_{\pi}^2}
+ \frac{\left({\bf p}^2\right)^2}{16 f_{\pi}^2} \right]^{1/2}
\left[ 1 + \frac{{\bf p}^2}{4f_{\pi}^2} \right]^{-1}
= 1 - \frac{{\bf p}^2}{2f_{\pi}^2}
+ \frac{\left({\bf p}^2\right)^2}{8 f_{\pi}^2} + \cdots \nonumber \\
&=& \cos(\phi/f_{\pi}) = 1 - \frac{\mbox{\boldmath $\phi$}^2}{2f_{\pi}^2}
+ \frac{\left(\mbox{\boldmath $\phi$}^2\right)^2}{24 f_{\pi}^2}
+ \cdots
\end{eqnarray}
To second order in the pion field all three representations
are the same.  Using the free field expression for the thermal
average of the field squared we get
\begin{equation}
\langle \sigma \rangle /f_{\pi} =
1 - \frac{N-1}{2} \left( \frac{T^2}{12 f_{\pi}^2}\right) + \cdots \, .
\end{equation}
For $N = 4$, the only value for which we can quantitatively compare
with QCD, this agrees with the result of Gasser and Leutwyler as
quoted in Eq. (\ref{qcon}); it was also derived in an independent
way by Eletsky \cite{Eletsky3}.

The coefficient of the term which is fourth order in the pion field
differs in sign and magnitude among the three representations.
It would be a miracle if the thermal average of $\sqrt{1 -
\mbox{\boldmath $\pi$}^2/f_{\pi}^2}, \,\, \cos(\phi/f_{\pi})$ and
the Weinberg expression were all the same!  But regarding the
order $(T^2/12f_{\pi}^2)^2$ we must recognize that the term
which is second order in the pion field gets modified due
to a one loop self-energy.  This was computed for each representation
in section 3.1 and the results listed in Eq. (\ref{self}).
The term fourth order in the pion field can be evaluated using
free fields.  The result is
\begin{equation}
\langle \left(\mbox{\boldmath $\phi$}^2\right)^2 \rangle
= \left(N^2 - 1\right) \left(\frac{T^2}{12}\right)^2 \, ,
\end{equation}
and is obviously representation independent.  The contributions
for each representation are
\begin{eqnarray}
\mbox{\boldmath $\pi$}: && 1 - \frac{N-1}{2} \left(\frac{T^2}
{12f_{\pi}^2} \right) \left[ 1 - \left( \frac{T^2}{12 f_{\pi}^2}
\right) \right]
- \frac{N^2-1}{8} \left( \frac{T^2}{12 f_{\pi}^2} \right)^2 \nonumber \\
{\bf p}: && 1 - \frac{N-1}{2} \left(\frac{T^2}
{12f_{\pi}^2} \right) \left[ 1 + \frac{N-1}{2}
\left( \frac{T^2}{12 f_{\pi}^2} \right)
\right] + \frac{N^2-1}{8} \left( \frac{T^2}{12 f_{\pi}^2} \right)^2
\nonumber \\
\mbox{\boldmath $\phi$}: && 1 - \frac{N-1}{2} \left(\frac{T^2}
{12f_{\pi}^2} \right)
\left[ 1 + \frac{N-2}{3} \left( \frac{T^2}{12 f_{\pi}^2} \right)
\right] + \frac{N^2-1}{24} \left( \frac{T^2}{12 f_{\pi}^2} \right)^2
\end{eqnarray}
where the second term in each line comes from the square of the
pion field and the last term comes from the pion field
in fourth order.  The sum of all terms is identical in all three
representations.
\begin{equation}
\langle \sigma \rangle /f_{\pi} = 1 - (N-1)
\left( \frac{T^2}{24 f_{\pi}^2}\right) -
\frac{(N-1)(N-3)}{2} \left( \frac{T^2}{24 f_{\pi}^2}\right)^2 + \cdots
\end{equation}
The miracle happens.  It is a consequence of the fact that physical
quantities must be independent of field redefinition.  What is more,
for $N$ = 4 it agrees with the result of Gasser and Leutwyler
quoted in Eq. (\ref{qcon}).  However, we emphasize once more that
this expression should not be used to infer a critical temperature
because it was derived under the assumption that the temperature
is small compared to $f_{\pi}$.

\section{Summary and Conclusion}

In this paper we have focussed on the linear and nonlinear versions
of the sigma model based on the group O($N$) at finite temperature.
Models of this kind are prototypes for physical theories, such as
QCD and electroweak theory.  Our main goal was to understand, both
conceptually and mathematically, whether the nonlinear model has
a symmetry restoring phase transition analogous to that of the linear
model.  We did show that the nonlinear model has a second order phase
transition by making use of the 1/$N$ expansion.  To leading and
subleading orders the critical temperature is even the same as
in the linear model.  (This cannot be true in general; eventually
there must be some dependence in the linear model on the value
of the quartic coupling $\lambda$.)  This expansion was facilitated by the
introduction of a Lagrange multiplier field.  In this way we could
see that there is a condensate at low temperature; this condensate
decreases in just the right way so as to conserve the constraint
on the field vector.  There is one particular temperature for which
the thermally averaged constraint is satisfied with no condensate
and with all excitations massless.  This is the critical
temperature.  We had to make a mathematical approximation at
the subleading order to get an analytical result.  We referred
to this as a ``high energy approximation".  It is directly
analogous to what one does in the mean field approximation to the
linear model.  It would be interesting to relax this approximation;
this is left as a future project.

Another goal was to carefully define and show how to compute
the ``pion decay constant" and the ``scalar quark condensate"
at finite temperature within the scope of these models.
The definitions also apply to full QCD but, of course, the
results will generally be different.  Only at very low temperatures
and for $N$ = 4 will the results be directly applicable to QCD
for the reasons discussed in the introduction.  Even within the
context of the sigma models, however, it would be interesting
to compute the next order correction to $f_{\pi}^2(T)$ at low
temperature.  It would also be interesting to compute
$f_{\pi}^2(T)$ near $T_c$.  These computations are now
underway.

In this paper we have not computed anything more complicated than
a one loop diagram.  Even the calculation of the ``scalar
quark condensate" to order $(T^2/12 f_{\pi}^2)^2$ only required
knowledge of one loop diagrams.  What other interesting physical
quantities can the reader compute in these models to one loop order?

Natural extensions of these models to better approximate full QCD
may be envisioned.   Following the philospophy of chiral perturbation
theory one may include higher derivative terms in the Lagrangian.
One may also add other mesonic and baryonic fields, especially
the vector mesons.  However, no matter how many extra terms are
added, one is still restricted from discussing the quark-gluon
plasma.

If the Higgs particle turns out to have an exceptionally large mass,
then a reasonable first approximation to the electroweak phase
transition might begin with a nonlinear version of the
Glashow--Weinberg--Salam model.  Gauging the nonlinear sigma model would
be a step in this direction.  This topic is also under investigation.

We hope to have stimulated the reader to make further progress
on these very interesting topics at finite temperature!

\section*{Acknowledgements}

We are grateful to A. Kovner and S. Jeon for discussions.
This work was supported by the U.S. Department of Energy under
grant number DE-FG02-87ER40328.

\newpage

\section*{Figure Captions}

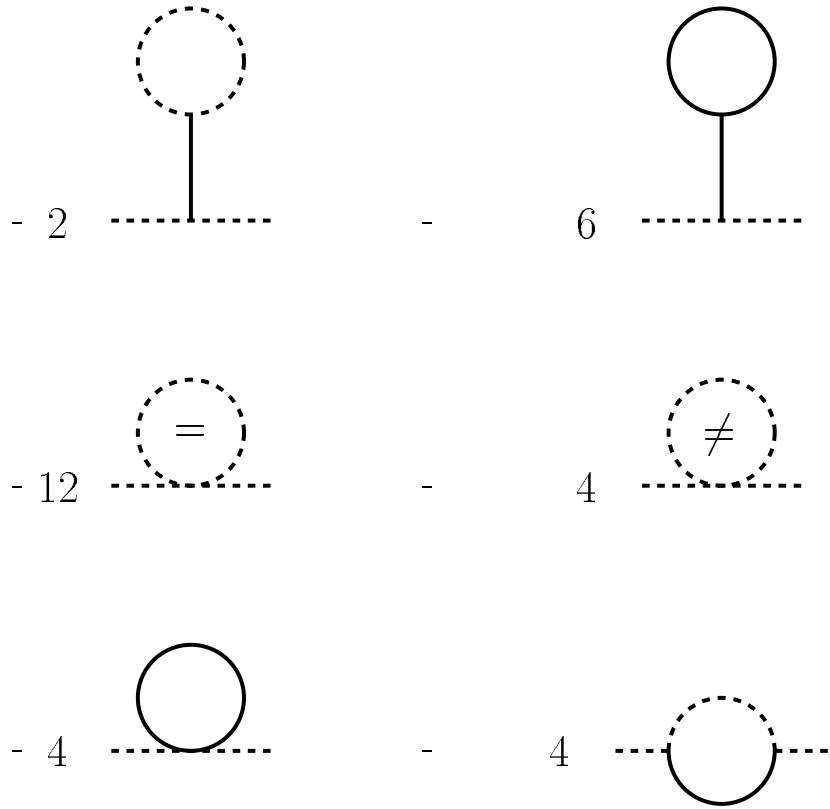
Figure 1: One loop self-energy diagrams for the pion in the linear
sigma model.  The dashed lines represent pions and the solid lines
represent sigma mesons.  The three-point vertices are $-\lambda v$
and the four-point vertices are $-\lambda/4$.  The $=$ and $\neq$
indicate that the pion in the loop has the same or different
quantum number than the external pion, respectively.  If $v$ is
fixed at its vacuum value of $f_{\pi}$ then the two tadpoles
contribute.  If $v$ is allowed to vary with temperature by maximizing
the pressure then the tadpoles are not to be included in the self-energy;
their effect is already included in the mean field mass via $v(T)$.\\

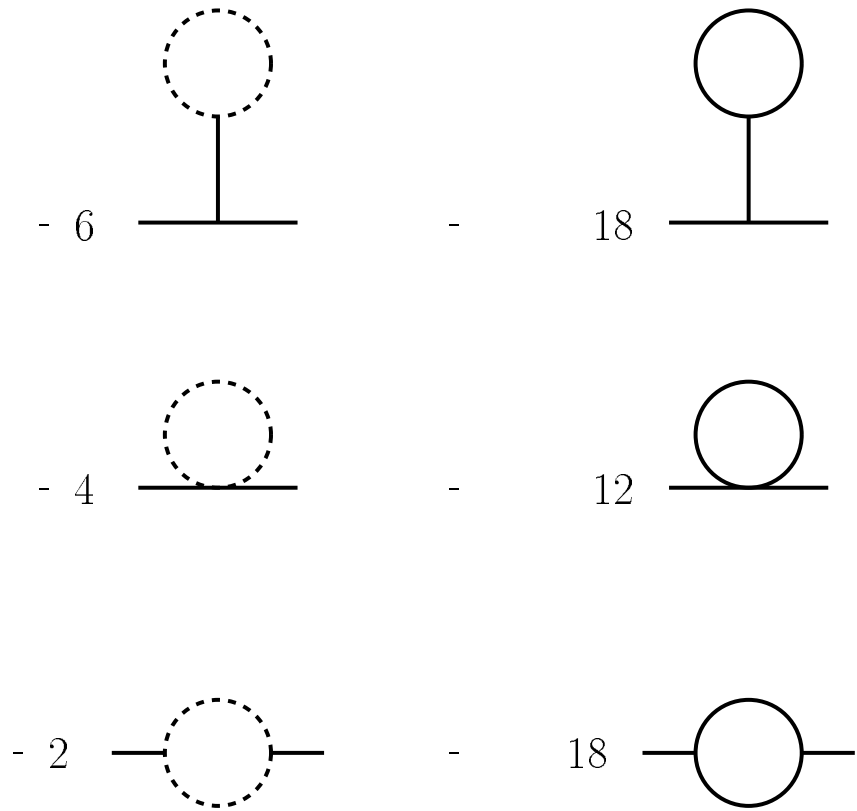
\noindent Figure 2: One loop self-energy diagrams for the sigma meson
in the linear sigma model.  See Figure 1 for remarks.\\

\noindent Figure 3: Contribution to the effective action at finite
temperature in the nonlinear sigma model corresponding to
Eq. (\ref{Seff}); the wavy lines represent the Lagrange
multiplier field $b$.\\

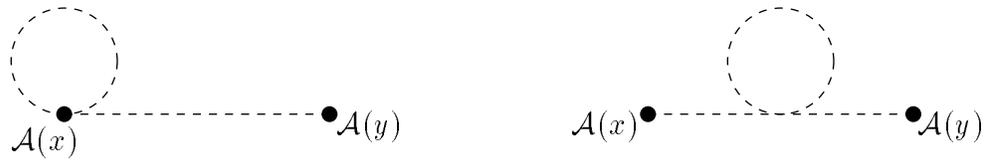
\noindent Figure 4: Vertex and self-energy contributions to the
axial vector correlation function in the nonlinear sigma model.\\

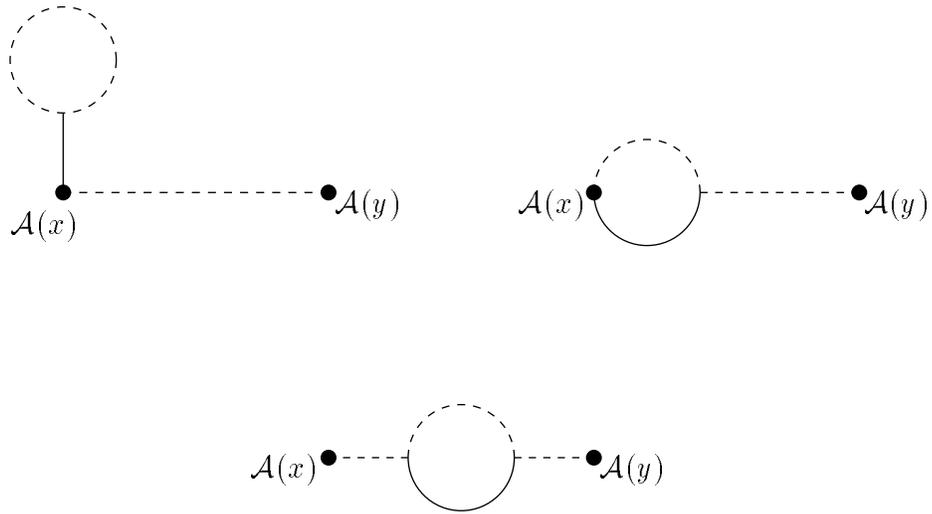
\noindent Figure 5: Vertex and self-energy contributions to the
axial vector correlation function in the linear sigma model.
See Figure 1 for remarks.

\end{document}